\documentclass[a4paper]{jpconf}
\usepackage{graphicx}

\usepackage{bbold}
\usepackage{graphicx,cite}
\usepackage{amsmath}
\usepackage{amsfonts}
\usepackage{amssymb}
\usepackage{rotating}

\usepackage{braket}

\usepackage{booktabs}

\usepackage{xcolor}
\usepackage{soul}

\begin{document}
\title{A principle for the Yukawa couplings}

\author{U J Salda\~na-Salazar}

\address{Facultad de Ciencias F\'isico-Matem\'aticas,
Benem\'erita Universidad Aut\'onoma de Puebla, \\
C.P. 72570, Puebla, Pue., M\'exico.}

\ead{ulisesjesus@protonmail.ch}

\begin{abstract}
Yukawa couplings in the Standard Model are introduced in its most general form, that  is, completely arbitrary complex numbers. Here we show that their origin could not be general
but dictated by a principle.
\end{abstract}

\section{Introduction}
There is a huge gap between knowing and not knowing. There are times though when their separation becomes  null and not knowing is knowing or vice-versa. 
Curiously, this is the current situation on the Yukawa couplings, which are introduced in order to parametrize the interactions between the two handedness of fermions and the Higgs field, 
with their inclusion being made by following the principle of generality. Therewith, we know
Yukawa couplings are, in general, complex and arbitrary numbers. But that is all we know about
them, i.e., that we do not know them. Here we try to briefly summarize the essence of a
recently proposed principle called the Flavor-Blind Principle \cite{Saldana-Salazar:2015raa} that gives an origin to the
Yukawa couplings in agreement to the emerged structures appearing 
under the phenomenological application of hierarchical fermion masses \cite{Hollik:2014jda}.

The same scalar field required to break spontaneously the electroweak symmetry can be 
employed to give masses in a gauge invariant and renormalizable way to all the fermions. 
This kind of interactions between the two handedness of fermions with the Higgs field are named Yukawa interactions. The Yukawa Lagrangian is normally written by demanding its most general form
\begin{eqnarray}
	-{\cal L}_Y =  \sum_{i,j=1}^3 \left(	\sum_{a=u,\nu} {\bf Y}^a_{ij} \bar{\psi}_{L,i}^a (i\sigma_2 \Phi^*) \psi_{R,j}^a +
	\sum_{b=d,e} {\bf Y}^b_{ij} \bar{\psi}_{L,i}^b \Phi \psi_{R,j}^b \right) + h.c.,
\end{eqnarray}
where ${\bf Y}^{a(b)}_{ij}$ are the Yukawa couplings which, due to generality, are arbitrary complex numbers. With only one scalar field, $\Phi$, the Yukawa matrices are proportional to the mass matrices. The constant producing such distinction appears  after the neutral component of the Higgs field acquires a non-zero vacuum expectation value, $v = 246$ GeV. 

It has become a well established fact, that matter in the subatomic world 
replicates itself thrice. That is, each of the four
types of fermions, whose distinction is given by their quantum numbers, 
appears three-fold; it is their mass the only one aspect allowing their differentiation. Each group of four is called family or generation, whereas, each one of the three members of a given fermion type is called flavor.

Phenomenologically, one finds that masses among the three flavors and on each fermion type
obey the hierarchical structure
\begin{eqnarray}
	m_1(M_Z) \ll m_2(M_Z) \ll m_3(M_Z).
\end{eqnarray}
Neutrinos, however, may be the exception; their squared mass differences still satisfy
a hierarchy 
\begin{eqnarray}
	\Delta m_{21}^2 \ll \Delta m_{31}^2,
\end{eqnarray}
but this is not necessarily the case for their individual masses.

Recently, by virtue of the aforementioned hierarchy, a mixing parametrization was built \cite{Hollik:2014jda} wherein its mixing parameters correspond to the four independent fermion mass ratios of the quark or lepton sector, respectively. The developed procedure 
found that fermion mixing originates from a very particular sequence of Yukawa matrices
\begin{eqnarray}
	{\bf M}_f = v {\bf Y}_f = v \left( {\bf Y}_{f,3} + {\bf Y}_{f,2} + {\bf Y}_{f,1} \right),
\end{eqnarray}
where
\begin{eqnarray} \label{Yuks}
	{\bf Y}_{f,3} = \begin{pmatrix}
		0 & 0 & 0 \\
		0 & 0 & 0 \\
		0 & 0 & y_3
	\end{pmatrix}, \qquad
	{\bf Y}_{f,2} = \begin{pmatrix}
		0 & 0 & 0 \\
		0 & 0 & y_2 \\
		0 & \pm y_2 & 0
	\end{pmatrix}, \quad \text{and} \quad
	{\bf Y}_{f,1} = \begin{pmatrix}
		0 & y_1 & {y}'_1 \\
		\pm y_{1} & 0 & 0 \\
		\pm{y}'_1 & 0 & 0
	\end{pmatrix}, 
\end{eqnarray}
with $y_3$ being real while $y_2$, $y_1$, and $y'_1$ being either real or purely imaginary. Also,
an unexplained feature was that  between these three matrices there must exist a hierarchy in correspondance to $m_{f,3} \gg m_{f,2} \gg m_{f,1}$.  It is our goal to find such an origin that gives precisely these Yukawa structures. 

This paper is organized with only two sections:  in Section \ref{FBP} we give account of the Flavor-Blind principle and briefly comment about all its implications while in Section \ref{sec:Conc} we conclude.

\section{The Flavor-Blind Principle} \label{FBP}
Consider the Standard Model (SM) Lagrangian\footnote{From this point onwards, we refer by SM a theory including massive neutrinos. Also, for the moment and due to simplicity, we take the massive nature of neutrinos to be of the Dirac type.} with its Yukawa part turned off. That is, we have not introduced yet the interactions between the Higgs field with the fermions. Also, assume we have an arbitrary number of families. Universality in the gauge interactions for $n$ families will imply an accidental global symmetry group given as
\begin{eqnarray}
	{\cal G}_F = U_L^Q(n) \times U_R^u(n) \times U_R^d(n) \times U_L^\ell(n) \times U_R^e(n) \times U_R^\nu(n).  
\end{eqnarray}
Basically, we cannot distinguish between any of the $n$ fermion families. The Flavor-Blind principle (FBP) then states that Yukawa interactions should be introduced as to conserve flavor-blindness. That is, the $n$ fermion fields should equally couple to the scalar field
\begin{eqnarray} \label{eq:n-permutation}
	{\bf Y}^{1\leftrightarrow 2 \leftrightarrow 3 \leftrightarrow
	 \cdots \leftrightarrow (n-1) \leftrightarrow n} = y_n \begin{pmatrix}
	 1 & 1 & 1 & \cdots & 1 \\
	 1 & 1 & 1 & \cdots & 1 \\
	 1 & 1 & 1 & \cdots & 1 \\
	\vdots & \vdots & \vdots & \vdots & \vdots \\
	1 & 1 & 1 & \cdots & 1 	 
	 \end{pmatrix}.
\end{eqnarray} 
The symmetry properties of this matrix are described by the permutational symmetry group $S_{nL}\otimes S_{nR}$. As this matrix is rank one then a change of basis will show that $(n-1)$ families are still massless while the remaining one is massive, $m_n = n y_n v/\sqrt{2}$. Notice that the same weak basis transformation applies to all fermions and it has the form
\begin{eqnarray} \label{WBT-Sn}
	{\bf O}_n = \begin{pmatrix}
\frac{1}{\sqrt{2}} & \frac{1}{\sqrt{3\cdot 2}} & \frac{1}{\sqrt{4\cdot 3}} & \cdots &
\frac{1}{\sqrt{n(n-1)}} &\frac{1}{\sqrt{n}} \\
-\frac{1}{\sqrt{2}} & \frac{1}{\sqrt{3\cdot 2}} & \frac{1}{\sqrt{4\cdot 3}} & \cdots &
\frac{1}{\sqrt{n(n-1)}} & \frac{1}{\sqrt{n}} \\
 0 & -\frac{2}{\sqrt{3\cdot 2}} & \frac{1}{\sqrt{4\cdot 3}} & \cdots &
\frac{1}{\sqrt{n(n-1)}} & \frac{1}{\sqrt{n}} \\
0 & 0 &  -\frac{3}{\sqrt{4\cdot 3}} & \cdots &
\frac{1}{\sqrt{n(n-1)}} & \frac{1}{\sqrt{n}} \\
\vdots & \vdots & \vdots & \cdots & \vdots & \vdots \\
0 & 0 & 0 & \cdots & \frac{1}{\sqrt{n(n-1)}} & \frac{1}{\sqrt{n}}\\
0 & 0 & 0 & \cdots & -\frac{n-1}{\sqrt{n(n-1)}} & \frac{1}{\sqrt{n}}
	\end{pmatrix}.
\end{eqnarray}
Interestingly, due to the generic form the Yukawa matrices acquire in the new basis
\begin{eqnarray} \label{eq:nnelelement}
	\widetilde{\bf Y}^{1\leftrightarrow 2 \leftrightarrow 3 \leftrightarrow
	 \cdots \leftrightarrow (n-1) \leftrightarrow n} = n y_n \begin{pmatrix}
	 0 & 0 & 0 & \cdots & 0 \\
	 0 & 0 & 0 & \cdots & 0 \\
	 0 & 0 & 0 & \cdots & 0 \\
	\vdots & \vdots & \vdots & \vdots & \vdots \\
	0 & 0  & 0 & \cdots & 1 	 
	 \end{pmatrix},
\end{eqnarray} 
we can easily see that our initial accidental and global symmetry group has broken into
\begin{eqnarray}
	{\cal G}_F \rightarrow U_L^Q(n-1) \times U_R^u(n-1) \times U_R^d(n-1) \times U_L^\ell(n-1) \times U_R^e(n-1) \times U_R^\nu(n-1),  
\end{eqnarray}
plus some additional $U(1)$ factors.

A new Yukawa matrix is now required in order to give mass to the still massless $(n-1)$ fermion fields. Application of the FBP means
\begin{eqnarray}
	{\bf Y}^{1\leftrightarrow 2 \leftrightarrow 3 \leftrightarrow
	 \cdots \leftrightarrow (n-1)} = \begin{pmatrix}
	 	\beta_{n-1} & \beta_{n-1} & \cdots & \beta_{n-1} & \alpha_{n-1} \\
	 	\beta_{n-1} & \beta_{n-1} & \cdots & \beta_{n-1} & \alpha_{n-1} \\
	 	\vdots & \vdots & \cdots & \vdots & \vdots \\
	 	\beta_{n-1} & \beta_{n-1} & \cdots & \beta_{n-1} & \alpha_{n-1} \\
	 	\alpha'_{n-1} & \alpha'_{n-1} &\cdots & \alpha'_{n-1} & \gamma_{n-1}
	 \end{pmatrix}.
\end{eqnarray}
The arbitrary parameters $\alpha_{n-1}$, $\beta_{n-1}$, $\alpha'_{n-1}$, and
$\gamma_{n-1}$ are, in general, complex parameters.
The symmetry properties of this matrix are described by the permutational symmetry group $S_{(n-1)L}\otimes S_{(n-1)R}$. It has rank two and using the same weak basis transformation of Eq.~\eqref{WBT-Sn} brings it into the form
\begin{eqnarray}
	\widetilde{\bf Y}^{1\leftrightarrow 2 \leftrightarrow 3 \leftrightarrow
	 \cdots \leftrightarrow (n-1)} = \begin{pmatrix}
	 0 & 0 & 0 & \cdots & 0 & 0& 0\\
	 0 & 0 & 0 & \cdots & 0 & 0& 0\\
	 0 & 0 & 0 & \cdots & 0 & 0 & 0\\
	\vdots & \vdots & \vdots & \vdots & \vdots & \vdots & \vdots\\
	0 & 0 & 0 & \cdots & 0 &0 & 0 \\
	0 & 0  & 0 & \cdots & 0 & \widetilde{\cal Y}_{n-1,n-1} &  	 \widetilde{\cal Y}_{n-1,n} \\
	0 & 0  & 0 & \cdots &0 & \widetilde{\cal Y}_{n,n-1} &  	 \widetilde{\cal Y}_{n,n}
	 \end{pmatrix},
\end{eqnarray}
where
\begin{eqnarray}
\widetilde{\cal Y}_{n-1,n-1} = 
	 \frac{n-1}{n}[\beta_{n-1} + \gamma_{n-1} - \alpha_{n-1}-\alpha'_{n-1} ],  \\
\widetilde{\cal Y}_{n-1,n} = 
	  \frac{\sqrt{n-1}}{n}[(n-1)(\beta_{n-1} - \alpha'_{n-1}) + \alpha_{n-1} - \gamma_{n-1}], \\
\widetilde{\cal Y}_{n,n-1} =  
	 \frac{\sqrt{n-1}}{n}[(n-1)(\beta_{n-1} - \alpha_{n-1}) + \alpha'_{n-1} - \gamma_{n-1}],  \\
\widetilde{\cal Y}_{n,n} = 
	  \frac{1}{n}\{ \gamma_{n-1} + (n-1)[(n-1)\beta_{n-1}+\alpha_{n-1}+\alpha'_{n-1}] \}.
\end{eqnarray}
We can now notice that besides the $n$-th family being massive the $(n-1)$-th one can also become massive after diagonalizing the complete $2 \times 2$ submatrix. We can see that our previous global symmetry group $U(n-1)^6$ has now
broken into $U(n-2)^6$ plus some additional $U(1)$ factors. 

In this way, by each step, we are providing mass to only one fermion family. When reaching
the $U(1)^6$ last step, i.e., with only one massless family, we need to consider that the properties of the needed Yukawa matrix should be $S_{2A}$ as shown in \cite{Saldana-Salazar:2015raa}. 

The repetition of this procedure leads to some important observations: 
\begin{enumerate}
	\item The accidental global symmetry group ${\cal G}_F$ needs to has the sequential breaking
	\begin{eqnarray}
		U(n)^6 \rightarrow U(n-1)^6 \rightarrow U(n-2)^6 \rightarrow \cdots \rightarrow U(2)^6 \rightarrow U(1)^6 \rightarrow U(1)_B \times U(1)_L,
	\end{eqnarray}
	with $B$ and $L$ being baryon and lepton number, respectively. When $n=3$
	this reminds us of minimal flavor violation which requires a similar breaking \cite{Barbieri:1995uv,
  Barbieri:1996ww, Barbieri:1997tu}.
	
	\item The introduced set of Yukawa matrices also has a sequential breaking of its
	symmetrical properties
	\begin{eqnarray}
		S_{nL}\otimes S_{nR} \rightarrow S_{(n-1)L}\otimes S_{(n-1)R} \rightarrow \cdots 
	\rightarrow S_{2L}\otimes S_{2R} \rightarrow S_{2A}.
	\end{eqnarray}
	\item Each step, after the weak basis transformation of Eq.~\eqref{WBT-Sn}, brings the following Yukawa structures
\begin{footnotesize}
	\begin{eqnarray}
		\begin{pmatrix}
			 & & & & &\\
			 & & & & &\\
			 & & & & &\\
			& & & & & \\
			& & & & & \times
		\end{pmatrix} \rightarrow
		\begin{pmatrix}
			 & & & & &\\
			 & & & & &\\
			 & & & & &\\
			& & & & \times & \times\\
			& & & & \times & \times
		\end{pmatrix} \rightarrow
		\begin{pmatrix}
			 & & & & &\\
			 & & & & &\\
			 & & & \times& \times& \times\\
			& & & \times& \times & \times\\
			& & & \times & \times & \times
		\end{pmatrix} \rightarrow \cdots \rightarrow
		\begin{pmatrix}
			\times &\times &\times &\times & \cdots &\times\\
			 \times &\times &\times &\times & \cdots &\times\\
			\times &\times &\times &\times & \cdots &\times\\
			\vdots &\vdots & \vdots &\vdots & \vdots & \vdots \\
			\times &\times & \times& \times & \times & \times
		\end{pmatrix} 
	\end{eqnarray}
\end{footnotesize}

		which basically are telling us how mixing phenomena occurrs. First, mixing in the $(n-1)$-$n$ sector with its initial individual mixing angle proportional to $m_{n-1}/m_n$. After it, nevertheless, we still require to introduce mixing in the same sector proportional to each of the new mass terms divided by $m_n$. Next, mixing in the $(n-2)$-$n$ sector happens. Then, mixing in $(n-3)$-$n$ and so on and so forth until reaching mixing between the first family with the last one. After all these rotations, the mass matrices become diagonal by parts with one $(n-1) \times (n-1)$ submatrix plus another $1\times 1$ one. Similarly, mixing between the $(n-1)$-th family with the rest happens and so on and so forth. This same observation was
		found when applying the mass hierarchy $m_3 \gg m_2 \gg m_1$ to the study of fermion
		mixing \cite{Hollik:2014jda}.
		
		\item Application of the FBP to the minimal case, that is, two families, naturally implies that individual mixing
		should happen satisfying a Gatto--Sartori--Tonin-like relation\footnote{The Gatto--Sartori--Tonin relation \cite{Gatto:1968ss} was the first expression to relate a mixing angle with a mass ratio as 
		\begin{eqnarray}
			\tan \theta_C \approx \sqrt{\frac{m_d}{m_s}} \approx 0.22,
		\end{eqnarray} which gives a very good approximation to the Cabibbo angle.} \cite{Saldana-Salazar:2015raa}
		\begin{eqnarray}
			\tan \theta_{12} = \sqrt{\frac{m_{1}}{m_2}}	.
		\end{eqnarray}
		
		\item Application of the FBP to the three family case gives a solvable system of linear equations with which we can easily reproduce the matrix structures of Eq.~\eqref{Yuks} \cite{Saldana-Salazar:2015raa}.
\end{enumerate}

\section{Conclusions}\label{sec:Conc}
We have seen how by demanding that massless fermion families equally couple to the Higgs field, therefore respecting their initial flavor-blindness in the gauge interactions,
gives us a series of phenomenological implications which are in agreement to the ones already found in the study of fermion mixing by virtue of hierarchical fermion masses \cite{Hollik:2014jda} and
minimal flavor violation \cite{Barbieri:1995uv,
  Barbieri:1996ww, Barbieri:1997tu}. 
This petition is called the Flavor-Blind Principle \cite{Saldana-Salazar:2015raa}.

Also, a very interesting remark about the usefulness of this approach is that it has recently been  seen to play an important role in solving the strong CP problem \cite{Diaz-Cruz:2016pmm}.

\ack
I am grateful to Wolfgang G. Hollik for a careful reading of this manuscript.
This work has been supported by CONACyT-Mexico under Contract No.~220498.

\end{document}